\documentclass[a4paper,11pt]{article}
\pdfoutput=1

\usepackage{jheppub}
\usepackage[T1]{fontenc}
\usepackage{amsmath,amssymb,amsthm}
\usepackage{tikz}
\usepackage{braket}
\usetikzlibrary{arrows.meta}

\newtheorem{theorem}{Theorem}
\newtheorem{lemma}[theorem]{Lemma}
\newtheorem{proposition}[theorem]{Proposition}
\theoremstyle{definition}

\newcommand{\A}{\mathcal A}
\newcommand{\Om}{\Omega}

\title{Proof of the entropic $c$-theorem}

\author{Jonah Kudler-Flam}

\affiliation{School of Natural Sciences, Institute for Advanced Study, Princeton, NJ, USA}

\emailAdd{jkudlerflam@ias.edu}

\abstract{We prove an entropic $c$-theorem in two dimensions by constructing a $c$-function from the mutual information between spacelike separated wedges, defined through the relative entropy. The proof follows Casini and Huerta's, with strong
subadditivity of entanglement entropy replaced by increasing differences of the mutual information, which we prove
directly in the continuum from the split property and standard properties of relative entropy. The benefit of relative entropy is that every quantity we use is finite and no cutoff ever needs to be introduced. The assumptions of the theorem are only the standard axioms of algebraic quantum field theory, the split property with finite
mutual information across a gap (both implied by modular $p$-nuclearity), and a finite
ultraviolet bound on the resulting $c$-functions. We also note that, formally, the $c$-function is proportional to the
minimal averaged null energy of a state that reproduces the vacuum correlations between the two wedges.}

\begin{document}
\maketitle
\flushbottom

\section{Introduction}

The $c$-theorem posits that there is a function, monotonically decreasing under renormalization group flow, that takes constant values at fixed points of the flow \cite{Zam86}. 
In the author's favorite paper, Casini and Huerta gave a proof of the $c$-theorem by combining strong
subadditivity of entanglement entropy and Lorentz invariance \cite{CH04,CH07}, and related entropic arguments were later given for the $F$-, $g$- and $a$-theorems \cite{CH12,CSLT16,CTT17}. The argument, however, is formal because entanglement
entropies of subregions in quantum field theory are ultraviolet divergent and so applying strong subadditivity requires a regulator.\footnote{The mutual information between separated regions was proposed as a
regulator-independent replacement in \cite{CHMY15}, and renormalization monotones intrinsic to the continuum limit were
constructed from recovery and scaling maps in \cite{Lashkari17}.}

In the algebraic formulation of quantum field theory, where the observables in a region form a von Neumann algebra, relative entropy can be defined directly in the continuum \cite{Araki76}. The mutual information, usually written as a linear combination of entanglement entropies, can be reformulated as a relative entropy and is finite for a causal diamond and a region at positive spacelike separation under nuclearity conditions \cite{PW22}. The purpose of this paper is to reformulate the ideas of Casini and Huerta using the mutual information in such a way that the entropic $c$-theorem can be proven directly in quantum field theory, without ever needing to introduce ultraviolet regulators at intermediate steps. The replacement for the strong subadditivity of entanglement entropy is the property of increasing differences of the mutual information, which we prove directly in the continuum. In turn, we will find independent left and right moving $c$-theorems, and note that, formally, these functions are proportional to the minimal null energy needed to reproduce the correlations between two spacelike separated Rindler wedges \cite{Wall17,CF20}.

In addition to the standard Haag--Kastler axioms \cite{HK64,Haag96} (locality, Poincar\'e covariance, positive energy, and a unique
vacuum), we assume that the algebras of nested causal diamonds satisfy the split property, with
finite vacuum mutual information across the gap between them. The split property provides the type I factors used in the
proof, and the finiteness makes the mutual information between a Rindler wedge and a causal diamond finite. Both follow
from modular $p$-nuclearity, a physically reasonable assumption on the high energy spectrum of the theory
\cite{BDL90a,BDL90b,PW22}. We also assume that the $c$-functions we define have finite ultraviolet values, which is the analog of $c_R^{\rm UV},c_L^{\rm UV}<\infty$. 

\section{Setup and formal argument}\label{sec:setup}

\begin{figure}[t]
\centering
\begin{tikzpicture}[scale=0.76,>=Stealth]
  \begin{scope}
  \clip (-2.65,-1.25) rectangle (2.65,3.0);
  \begin{scope}[rotate=45]
    \def\a{1.25}\def\b{1.6}\def\L{6}
    \fill[gray!22] (-\L,\a) rectangle (0,\a+\L);
    \fill[gray!22] (\b,-\L) rectangle (\b+\L,0);
    \draw[gray] (-\L,\a) -- (0,\a) -- (0,\a+\L);
    \draw[gray] (\b,-\L) -- (\b,0) -- (\b+\L,0);
    \fill[yellow!35] (0,0) rectangle (\b,\a);
    \draw (0,0) rectangle (\b,\a);
    \draw[dashed] (0,\a) -- (\b,0);
    \draw[gray,thin] (0,0) -- (-0.3,0) (0,0) -- (0,-0.3);
    \draw[<->,thin] (-0.22,0) -- (-0.22,\a) node[midway,below left=-2pt] {\scriptsize $a$};
    \draw[<->,thin] (0,-0.22) -- (\b,-0.22) node[midway,below right=-2pt] {\scriptsize $b$};
    \fill (0,0) circle (1.4pt) node[below=1pt] {$p$};
    \node at (0.62*\b,0.62*\a) {\small $r$};
  \end{scope}
  \end{scope}
  \node at (-1.9,0.95) {$L_a$};
  \node at (1.75,1.0) {$R_b$};
  \node at (0.15,0.55) {\small $D$};
  \node at (0,-1.55) {(a)};
\end{tikzpicture}\hspace{2pt}
\begin{tikzpicture}[scale=0.76,>=Stealth]
  \begin{scope}
  \clip (-2.65,-1.25) rectangle (2.65,3.0);
  \begin{scope}[rotate=45]
    \def\a{0.95}\def\A{1.85}\def\b{0.95}\def\B{1.85}\def\L{6}
    \fill[gray!22] (-\L,\A) rectangle (0,\A+\L);
    \fill[gray!22] (\B,-\L) rectangle (\B+\L,0);
    \draw[gray] (-\L,\a) -- (0,\a) -- (0,\a+\L);
    \draw[gray] (-\L,\A) -- (0,\A);
    \draw[gray] (\b,-\L) -- (\b,0) -- (\B+\L,0);
    \draw[gray] (\B,-\L) -- (\B,0);
    \fill[yellow!35] (0,0) rectangle (\b,\a);
    \draw (0,0) rectangle (\b,\a);
    \draw[gray,thin] (0,0) -- (-0.72,0) (0,0) -- (0,-0.72);
    \draw[<->,thin] (-0.25,0) -- (-0.25,\a) node[midway,below left=-3pt] {\scriptsize $a$};
    \draw[<->,thin] (-0.62,0) -- (-0.62,\A) node[pos=0.62,below left=-3pt] {\scriptsize $A$};
    \draw[<->,thin] (0,-0.25) -- (\b,-0.25) node[midway,below right=-3pt] {\scriptsize $b$};
    \draw[<->,thin] (0,-0.62) -- (\B,-0.62) node[pos=0.62,below right=-3pt] {\scriptsize $B$};
    \fill (0,0) circle (1.4pt) node[below=1pt] {$p$};
  \end{scope}
  \end{scope}
  \node at (-2.1,1.6) {$L_A$};
  \node at (2.1,1.6) {$R_B$};
  \node at (0,-1.55) {(b)};
\end{tikzpicture}\hspace{2pt}
\begin{tikzpicture}[scale=0.76,>=Stealth]
  \begin{scope}
  \clip (-2.65,-1.25) rectangle (2.65,3.0);
  \begin{scope}[rotate=45]
    \def\a{1.0}\def\b{1.0}\def\L{6}
    \fill[gray!22] (-\L,\a) rectangle (0,\a+\L);
    \draw[gray] (-\L,\a) -- (0,\a) -- (0,\a+\L);
    \draw[gray,dashed] (\b,-\L) -- (\b,0) -- (\b+\L,0);
    \fill[red!25] (\b+0.3,-0.85) rectangle (\b+1.1,-0.2);
    \draw[red!60!black] (\b+0.3,-0.85) rectangle (\b+1.1,-0.2);
    \fill[yellow!35] (0,0) rectangle (\b,\a);
    \draw (0,0) rectangle (\b,\a);
    \draw[densely dotted] (\b-0.15,-1.0) rectangle (\b+1.3,0.45);
    \fill (0,0) circle (1.4pt) node[below] {$p$};
  \end{scope}
  \end{scope}
  \node at (-1.9,1.05) {$L_a$};
  \node at (1.57,0.83) {\small $Y$};
  \node at (1.62,2.05) {\small $\hat D$};
  \node at (0,-1.55) {(c)};
\end{tikzpicture}
\caption{(a) The diamond $D=D(a,b)$ with corner $p$ and the Rindler wedges $L_a$ and $R_b$ in its causal complement. (b) The null-translated wedges $L_A\subset L_a$ and $R_B\subset R_b$ (shaded). (c) A causal diamond $Y\subset R_b$ and a larger causal diamond $\hat D$ containing its closure that is still spacelike to $L_a$.}
\label{fig:regions}
\end{figure}

We use light-cone coordinates $u=t-x$, $v=t+x$. Fixing a point, $p$, at the origin and $a,b>0$, the diamond $D(a,b)=\{0<u<a,\ 0<v<b\}$ has a spacelike
diagonal of proper length $r=\sqrt{ab}$, and its causal complement is the union of two Rindler wedges (see figure~\ref{fig:regions}a)
\begin{align}
L_a=\{u>a,\ v<0\},\qquad R_b=\{u<0,\ v>b\}.
\end{align}
We write $\A(O)$ for the algebra of observables in a region $O$, $|\Om\rangle$ for the vacuum, $S_{\A}(\psi\|\varphi)$ for
the relative entropy of two states on an algebra $\A$, and
\begin{align}
I(O_1:O_2)=S_{\A(O_1)\vee\A(O_2)}\big(\Om\,\big\|\,\Om_{O_1}\otimes\Om_{O_2}\big)
\end{align}
for the mutual information of two spacelike separated regions. This is well-defined when the two regions are split
\cite{DL84}, which effectively factorizes the Hilbert space as $\mathcal H=\mathcal H_1\otimes\mathcal H_2$, with $\A(O_1)$ acting on the
first factor and $\A(O_2)$ on the second. Equivalently,
the algebra $\mathcal N=B(\mathcal H_1)\otimes1$ of all operators on the first factor, a type I factor, lies between
$\A(O_1)$ and the commutant $\A(O_2)'$. Physically, such a factorization is expected across any finite corridor
separating the two regions. Up to infrared issues that we will ameliorate shortly, the $c$-function is given by the rate at
which the mutual information of the two wedges falls as the diamond grows,
\begin{align}\label{eq:C}
C(r)=-6a\,\frac{\partial}{\partial a}\,I(L_a:R_b),\qquad r=\sqrt{ab}.
\end{align}
Boost invariance implies that $I(L_a:R_b)$ only depends on $r$, so $C(r)=-3r\,\partial_rI(L_a:R_b)$.

Before proving the $c$-theorem rigorously, we provide a formal argument by treating the regions as if they had finite entanglement entropies. In terms of entropies, the mutual information is $I(L_a:R_b)=2S_W-S(r)$,
where $S_W$ is the entropy of a single wedge, which does not depend on the position of the corner of the wedge, and $S(r)$ is the entropy of a diamond of
diagonal $r$, which is equal to $S(L_a\cup R_b)$ by the purity of the vacuum state. Then, $C(r)=3rS'(r)$, which is three times the entropic $c$-function of Casini and Huerta. In conformal field theories, $S(r)=\frac c3\log r+\text{const}$ \cite{HLW94,CC04}, so $C=c$.

Let $a \leq A$ and $b\leq B$, which define new left and right wedges $L_A$ and $R_B$ that are null translations of $L_a$ and $R_b$. In particular, these are proper subregions of $L_a$ and $R_b$ (see figure \ref{fig:regions}b). Strong subadditivity of entanglement entropy \cite{LR73} implies that
\begin{align}
I(L_a:R_b)-I(L_A:R_b)&\ \ge\ I(L_a:R_B)-I(L_A:R_B),\label{eq:sm}\\
r\,\frac{dC}{dr}=2b\,\partial_bC&=-12ab\,\partial_a\partial_bI(L_a:R_b)\le0,\nonumber
\end{align}
where the second line is the infinitesimal form of the first, $\partial_a\partial_bI(L_a:R_b)\ge0$, and uses that $C$
depends only on $r$. This proves the monotonicity of $C$.

While in intermediate steps we used the entanglement entropy and its strong subadditivity, ultimately the crucial inequality is \eqref{eq:sm}, which only references mutual information. This motivates proving an inequality like \eqref{eq:sm} directly in the continuum theory, where the entanglement entropies do not exist but the mutual informations do. This is the goal of the remainder of the paper. 

\section{The \texorpdfstring{$\boldsymbol{c}$}{c}-theorem}\label{sec:theorem}

\subsection{Statement}\label{sec:statement}

Even though $L_a$ and $R_b$ have finite separation so there should be no ultraviolet divergence in the mutual information, there can in general be an infrared divergence due to their infinite extents. Fortunately, the inequality \eqref{eq:sm} only depends on the differences $I(L_a:R_b)- I(L_A:R_b)$ and $I(L_a:R_B)- I(L_A:R_B)$ which should be better behaved. To tame the potential infrared divergence, we define two new functions that are designed to capture these differences (see figure~\ref{fig:regions}c)
\begin{align}\label{eq:Dpm}
\begin{aligned}
\mathcal D_R(a,A;b)&=\sup_{Y\subset R_b}\big[I(L_a:Y)-I(L_A:Y)\big],\\
\mathcal D_L(a;b,B)&=\sup_{X\subset L_a}\big[I(X:R_b)-I(X:R_B)\big],
\end{aligned}
\end{align}
with the suprema over causal diamonds, for $a\le A$ and $b\le B$ (for $A<a$ we set $\mathcal D_R(a,A;b)=-\mathcal D_R(A,a;b)$,
and similarly for $\mathcal D_L$). Both functions are non-negative due to the monotonicity of relative entropy \cite{Uhlmann77} because $L_A\subset L_a$
and $R_B\subset R_b$. From here, we can define the one-sided $c$-functions
\begin{align}\label{eq:Cpm}
C_R(a,b)=6a\,\frac{\partial}{\partial A}\mathcal D_R(a,A;b)\Big|_{A=a},\qquad
C_L(a,b)=6b\,\frac{\partial}{\partial B}\mathcal D_L(a;b,B)\Big|_{B=b},
\end{align}
with derivatives from the right. $C_R$ translates $L_a$ along $u$ and so probes the right-movers, while $C_L$ translates $R_b$ along $v$ and probes the left-movers.
Boost invariance makes these depend only on $r=\sqrt{ab}$, so we will write $C_{R,L}(r)$. 

The first assumption is that the vacuum has finite mutual information across any gap for bounded regions. That is, for a causal diamond $D$
whose closure lies in a larger causal diamond $\tilde D$, the inclusion $\A(D)\subset\A(\tilde D)$ is split and
\begin{align}\label{eq:finiteI}
I\big(\A(D):\A(\tilde D)'\big)<\infty.
\end{align}
By monotonicity of relative entropy, $D$ then has finite mutual information with any region spacelike separated from
$\tilde D$. We also take the algebra of each causal diamond to be generated by those of the causal diamonds whose closures
it contains. If the algebras of causal diamonds are factors, the first assumption follows from modular
$p$-nuclearity.\footnote{Modular $p$-nuclearity means that the map
$\Xi:x\mapsto\Delta_{\A(\tilde D)}^{1/4}x\ket{\Omega}$ from $\A(D)$ to the Hilbert space,
where $\Delta_{\A(\tilde D)}$ is the modular operator of $\A(\tilde D)$ for the vacuum, is $p$-nuclear. For an operator between Hilbert spaces, this would be the condition $\operatorname{Tr}|\Xi|^p<\infty$ with $0< p< 1$.
$p$-nuclearity implies the split property
\cite{BDL90a,BDL90b}. The theorem of \cite{PW22} giving finite mutual information assumes hyperfiniteness of the
algebras, which holds because the algebra of a causal diamond is generated by an increasing
sequence of type I factors between smaller diamonds, hence by matrix algebras \cite{BDF87}, and its commutant
$\A(\tilde D)'=J\A(\tilde D)J$, with $J$ the modular conjugation of the vacuum, is isomorphic to $\A(\tilde D)$ and so is also hyperfinite.}

The second assumption is that the one-sided $c$-functions have finite ultraviolet values. Namely, there are constants $c_R^{\rm UV},c_L^{\rm UV}<\infty$ such that $\mathcal D_R(a,A;b)\le\frac{c_R^{\rm UV}}6\log\frac Aa$ and $\mathcal D_L(a;b,B)\le\frac{c_L^{\rm UV}}6\log\frac Bb$ for $a\le A$ and $b\le B$.

\begin{theorem}\label{thm:main}
Under the Haag--Kastler axioms and the two assumptions above, $C_R(a,b)$ and $C_L(a,b)$ depend only on $r=\sqrt{ab}$, and both are non-negative and non-increasing in $r$.
\end{theorem}

The proof rests on one inequality, which replaces the strong subadditivity step of section~\ref{sec:setup}.

\begin{proposition}\label{P:Ymono}
Under the Haag--Kastler axioms and the first assumption above, for $a\le A$ and causal diamonds $Y_1\subset Y_2\subset R_b$,
\begin{align}\label{eq:Ymono}
I(L_a:Y_1)-I(L_A:Y_1)\ \le\ I(L_a:Y_2)-I(L_A:Y_2).
\end{align}
\end{proposition}

\noindent This is the statement that the mutual information has increasing differences. The loss of correlations $I(L_a:Y)-I(L_A:Y)$ from shrinking $L_a$ to $L_A$ is non-decreasing in $Y$. If we formally replace the causal diamonds with the wedges $R_B\subset R_b$, \eqref{eq:Ymono} becomes \eqref{eq:sm}.

\subsection{Increasing differences}\label{sec:SSA}

For any causal
diamond $Y\subset R_b$, there is a causal diamond
$\hat D$ containing the closure of $Y$ that is still spacelike to all of $L_a$ (figure~\ref{fig:regions}c). By
assumption, the pair of algebras $(\A(\hat D)',\A(Y))$ is split, and because $\A(L_a)\subset\A(\hat D)'$, $(\A(L_a),\A(Y))$ is also
split. Monotonicity of relative entropy and \eqref{eq:finiteI} then give
\begin{align}
I(L_a:Y)\le I\big(\A(Y):\A(\hat D)'\big)<\infty.
\end{align}
The same holds for causal diamonds $X\subset L_a$ paired with $R_b$.

We prove \eqref{eq:Ymono} by approximating $\A(Y_1)$ from inside with type I factors. Besides monotonicity, we use three standard properties of mutual information \cite{OP93}, where we write $I(L_a:\mathcal M)$ for the
mutual information of $\A(L_a)$ and an algebra $\mathcal M$. (1) If commuting algebras $\mathcal X$, $\mathcal Y$ and
$\mathcal Z$ generate their tensor product, $\mathcal{X}\vee \mathcal{Y} \vee \mathcal{Z} =\mathcal{X}\otimes \mathcal{Y} \otimes \mathcal{Z} $, then
\begin{align}\label{eq:chain}
I(\mathcal X:\mathcal Y\vee\mathcal Z)+I(\mathcal Y:\mathcal Z)=I(\mathcal X\vee\mathcal Y:\mathcal Z)+I(\mathcal X:\mathcal Y),
\end{align}
where either side may be infinite. By the chain rule for relative entropy, both sides equal the relative entropy of the vacuum state with respect to the tensor product of its reduced states on the three algebras. (2) If we consider increasing sequences of algebras, $\{ \mathcal{X}_n\}$ and $\{ \mathcal{Y}_n\}$, the mutual information between the algebras is increasing and converges to the mutual information of the algebras they generate \cite{Kosaki86}. (3) The
mutual information of any algebra with a $d\times d$ matrix algebra is bounded from above by $2\log d$ 
\cite{PP86}.

\begin{lemma}\label{L:typeI}
Let $Y\subset R_b$ be a causal diamond and $\mathcal N\subset\A(Y)$ a type I factor. Then
\begin{align}
I(L_a:\mathcal N)-I(L_A:\mathcal N)\ \le\ I(L_a:Y)-I(L_A:Y).
\end{align}
\end{lemma}

\begin{proof}
We first let $\mathcal N$ be a type I$_d$ factor. Then $\A(Y)=\mathcal N\otimes\mathcal G$ with $\mathcal G=\mathcal N'\cap\A(Y)$. Since $(\A(L_a),\A(Y))$ is split, $\A(L_a)\vee\mathcal N\vee\mathcal G=\A(L_a)\otimes\mathcal N\otimes\mathcal G$, and likewise for $L_A\subset L_a$, so the chain rule
\eqref{eq:chain} for these algebras reads
\begin{align}
\begin{aligned}
    &I(L_a:Y)+I(\mathcal N:\mathcal G)=I\big(\A(L_a)\vee\mathcal N:\mathcal G\big)+I(L_a:\mathcal N),
\\
    &I(L_A:Y)+I(\mathcal N:\mathcal G)=I\big(\A(L_A)\vee\mathcal N:\mathcal G\big)+I(L_A:\mathcal N).
\end{aligned}
\end{align}
Every term is finite, since each left side is at most $I(L_a:Y)+2\log d$ by (3), and subtracting the two
identities gives
\begin{align}
\begin{aligned}
&\big[I(L_a:Y)-I(L_A:Y)\big]-\big[I(L_a:\mathcal N)-I(L_A:\mathcal N)\big]\\
&\qquad=I\big(\A(L_a)\vee\mathcal N:\mathcal G\big)-I\big(\A(L_A)\vee\mathcal N:\mathcal G\big)\ \ge\ 0
\end{aligned}
\end{align}
where the inequality follows from monotonicity, since $\A(L_A)\subset\A(L_a)$. A general type I factor $\mathcal N$ is generated by an increasing
sequence of unital matrix subfactors. The lemma then holds for each of them, and their mutual informations with $L_a$ and $L_A$ converge from below to the mutual informations with $\mathcal N$, which are finite because they are at most $I(L_a:Y)$.
\end{proof}

To prove Proposition~\ref{P:Ymono}, let $D_n$ be causal diamonds increasing to $Y_1$, with the closure of each in the
next, and $\mathcal N_n$ type I factors with $\A(D_n)\subset\mathcal N_n\subset\A(D_{n+1})$, which exist because these inclusions are split by assumption \cite{BDF87}. The $\mathcal N_n$ increase
and generate $\A(Y_1)$, because the $\A(D_n)$ do, so the mutual informations of $L_a$ and $L_A$ with $\mathcal N_n$
increase to those with $Y_1$. Lemma~\ref{L:typeI} with $\mathcal N=\mathcal N_n$ and $Y=Y_2$ then gives \eqref{eq:Ymono}.

\subsection{Proof of Theorem~\ref{thm:main}}\label{sec:proofs}

We are now in a position to prove the $c$-theorem. We treat the right movers, and the left movers are treated
identically. The function $\mathcal D_R$ has three crucial properties. First, it is additive,
\begin{align}\label{eq:add}
\mathcal D_R(a,A;b)+\mathcal D_R(A,A';b)=\mathcal D_R(a,A';b),
\end{align}
because by Proposition~\ref{P:Ymono} the loss of correlations $I(L_a:Y)-I(L_A:Y)$ grows with $Y$, so each supremum is a limit along the
causal diamonds in $R_b$ ordered by inclusion, and the limit of a sum is the sum of the
limits. Second, a boost about $p$ maps $L_a$ to $L_{\mu a}$ and the causal diamonds in $R_b$ to those in $R_{b/\mu}$, so
\begin{align}\label{eq:boost}
\mathcal D_R(\mu a,\mu A;b/\mu)=\mathcal D_R(a,A;b).
\end{align}
Third, for $b\le B$ the causal diamonds in $R_B$ are always in $R_b$ as well, so for $a\le A$
\begin{align}\label{eq:bmono}
\mathcal D_R(a,A;b)\ge\mathcal D_R(a,A;B).
\end{align}

Additivity and boost invariance reduce $\mathcal D_R$ to a function of one variable. Taking $\mu=b$ in \eqref{eq:boost}
gives $\mathcal D_R(a,A;b)=\mathcal D_R(ab,Ab;1)$. With $F(x)=\mathcal D_R(1,e^x;1)$,
\eqref{eq:add} gives $\mathcal D_R(ab,Ab;1)=\mathcal D_R(ab,1;1)+\mathcal D_R(1,Ab;1)$, that is,
\begin{align}\label{eq:F}
\mathcal D_R(a,A;b)=F(\log Ab)-F(\log ab).
\end{align}
The function $F$ is finite by the
ultraviolet bound and non-decreasing because $\mathcal D_R\ge0$. With $x=\log ab$ and $h=\log(A/a)\ge0$, \eqref{eq:F} reads
$\mathcal D_R(a,A;b)=F(x+h)-F(x)$. Increasing $b$ at fixed $a$ and $A$ increases $x$ at fixed $h$, so by
\eqref{eq:bmono} the increment $F(x+h)-F(x)$ is non-increasing in $x$. A non-decreasing function whose increments decrease is concave \cite{BD15}, so $F$ has a right derivative
$F'$, and since $a\partial_A$ at $A=a$ is $\partial_h$ at $h=0$, \eqref{eq:Cpm} gives
\begin{align}
C_R(a,b)=6F'(\log ab).
\end{align}
Therefore $C_R$ depends only on $r=\sqrt{ab}$, it is non-negative because $F$ is non-decreasing, and it is non-increasing
in $r$ because $F$ is concave. Its ultraviolet limit ($r\to0$) exists and is at most $c_R^{\rm UV}$, because the
ultraviolet bound reads $F(x+h)-F(x)\le c_R^{\rm UV}h/6$.

The one-sided $c$-functions also have a simple interpretation in terms of null energy. Let
$E_{\min}(a,Y)$ be the infimum of the averaged null energy $\int T_{uu}\,du$ of the first factor, over all vectors in $\mathcal H\otimes\mathcal H$ that have the same correlation functions as the vacuum for operators in $L_a$ (acting on the first factor) and $Y$ (acting on the second). The ant formula, conjectured by Wall \cite{Wall17} and proved by Ceyhan and
Faulkner \cite{CF20} using half-sided modular inclusions \cite{Borchers92,Wiesbrock93} (see also \cite{HL25}), gives $-\partial_aI(L_a:Y)=2\pi E_{\min}(a,Y)$, and hence, formally,
\begin{align}
C_R(\sqrt{ab})=12\pi a\lim_YE_{\min}(a,Y).
\end{align} 
Thus $C_R$ measures the null energy it costs to reproduce the correlations between the two
wedges. A different connection between averaged null energy and renormalization group monotonicity was found in \cite{HM23a,HM23b}. 

\section{Fixed points}\label{sec:examples}

The limits $C_{R,L}(0^+)$ and $C_{R,L}(\infty)$ exist by monotonicity and are finite, and they play the role of entropic central charges of the ultraviolet and infrared theories, although we do not prove that they equal the values of the limiting fixed-point theories.
At fixed points, dilations act by a unitary group that fixes the vacuum, and a
dilation by $\lambda$ maps $L_a$, $R_b$ and the causal diamonds in $R_b$ to $L_{\lambda a}$, $R_{\lambda b}$ and the causal
diamonds in $R_{\lambda b}$, so
\begin{align}
\mathcal D_R(\lambda a,\lambda A;\lambda b)=\mathcal D_R(a,A;b),\quad C_R(\lambda a,\lambda b)=C_R(a,b),
\end{align}
and because $C_R$ depends only on $ab$ it is constant. 

The final consideration is whether our entropic central charges agree with the conventional central charges defined via the stress tensor, whenever defined. First, we may consider a conformal field theory whose local observables factorize into left- and right-moving chiral algebras, $\mathcal H=\mathcal H_R\otimes\mathcal H_L$ and $\A(I\times J)=\A_R(I)\otimes \A_L(J)$
for a $u$-interval $I$ and a $v$-interval $J$, where $\A_R$ and $\A_L$ have central charges $c_R$ and $c_L$. For these theories, the mutual information for $Y=\{-\ell<u<0,\ b<v<b+m\}$ splits as
\begin{align}
I(L_a:Y)=I_R(\eta_u)+I_L(\eta_v),
\end{align}
with conformal cross ratios $\eta_u=\ell/(a+\ell)$ and $\eta_v=m/(b+m)$, where $I_{R,L}(\eta)$ is the two-interval mutual
information of the chiral theory. We regard $I_R$ as a function of $x=\log\frac\eta{1-\eta}$, which is $\log\frac\ell a$
for $\eta=\eta_u$. With $h=\log(A/a)$, replacing $L_a$ by $L_A$ lowers $x$ by $h$, so the loss of correlations
$I(L_a:Y)-I(L_A:Y)$ is the increment of $I_R$ from $x-h$ to $x$. By \eqref{eq:Ymono} it grows with $\ell$, and so with $x$.
Therefore, $I_R$ is non-decreasing in $x$ and has increasing increments, so it is convex in $x$ \cite{BD15}, and the
coefficient of its ultraviolet logarithm
\begin{align}
\kappa_R=\lim_{\eta\to1}\frac{I_R(\eta)}{\log\frac1{1-\eta}}
\end{align}
exists. As $Y$ grows to fill $R_b$, $x\to\infty$ and this loss approaches $\kappa_Rh$, so $\mathcal D_R(a,A;b)=\kappa_R\log\frac Aa$ and
$C_R=6\kappa_R$, and likewise $C_L=6\kappa_L$. 

The relation $C_{R,L}=6\kappa_{R,L}$, with $\kappa_{R,L}$ computed from the chiral algebras, holds for any rational conformal field theory that can be viewed as a finite-index extension
$\A\supset \A_R\otimes \A_L$ \cite{LR95,KLM01} of the product of its left- and right-moving chiral algebras. There, for two spacelike separated regions $O_1$ and $O_2$, let $E=E_{O_1}\otimes E_{O_2}$ be the product of the
projections onto the chiral subtheory $\A_R\otimes\A_L$ in each region, which preserve the vacuum and project out the
non-chiral primaries (local conditional expectations). This
preserves the product state $\Om_{O_1}\otimes\Om_{O_2}$, but not the vacuum of the joint non-chiral theory, so the chain rule for
relative entropy together with the Pimsner--Popa inequality gives \cite{PP86,OP93,LX18}
\begin{align}
0\le I_{\A}(O_1:O_2)-I_{\A_R\otimes\A_L}(O_1:O_2)=S_{\A(O_1)\vee\A(O_2)}(\Om\,\|\,\Om\circ E)\le2\log\iota,
\end{align}
with $\iota$ the (finite) index of each local inclusion. Since $C_R$ is constant at a fixed point, the concave function $F$ of \eqref{eq:F} has constant right derivative and is
therefore linear, so $\mathcal D_R(a,A;b)=\frac{C_R}6\log\frac Aa$, both for the full theory and for its chiral subtheory.
By the bound above, their $\mathcal D_R$ differ by at most $2\log\iota$ for every $A/a$, so the two coefficients agree,
and $C_{R,L}=6\kappa_{R,L}$ for every such theory.

The remaining step, $\kappa_R=c_R/6$ and $\kappa_L=c_L/6$, is what non-rigorous twist-field calculations universally give for the leading logarithm of the
two-interval mutual information as the intervals approach each other \cite{CC04,CCT09}. For rational conformal field theories, this follows from a conjecture of Xu on the vacuum
mutual information of two intervals, which has been proved for a subclass of theories \cite{Xu20}. It would be interesting
to prove this for general rational and irrational conformal field theories.

\acknowledgments
I am grateful for discussions with Nima Lashkari, Jonathan Sorce, and Edward Witten.
I acknowledge support from the Institute for Advanced Study and the National
Science Foundation under Grant PHY-2514611. I used the assistance of Claude and ChatGPT for development of arguments and proofreading. All derivations were carried out and written up by the author, who takes full
responsibility for the claims.

\bibliographystyle{JHEP}
\bibliography{main}

\end{document}